\documentclass[runningheads]{llncs}
\usepackage[T1]{fontenc}

\usepackage{amsmath, amssymb, amsfonts}
\usepackage{float}
\usepackage[pdftex]{graphicx}
\usepackage{caption}
\usepackage{subcaption}
\usepackage{hyperref}
\usepackage{circledsteps}

\title{Evaluating PDE discovery methods for multiscale modeling of biological signals}
\titlerunning{Evaluating PDE discovery methods for multiscale modeling}
\author{Andr\'ea Ducos\orcidID{0009-0006-3711-4150} \and
Audrey Denizot\orcidID{0000-0002-3336-0163} \and
Thomas Guyet\orcidID{0000-0002-4909-5843}\and
Hugues Berry\orcidID{0000-0003-3470-683X}}
\authorrunning{A. Ducos et al.}
\institute{AIstroSight, Inria, Hospices Civils de Lyon, Université Claude Bernard Lyon 1, \\
56 Bld Niels Bohr, Villeurbanne, 69603, France\\
\email{andrea.ducos@inria.fr}
}

\begin{document}
\emergencystretch 3em

\maketitle 

\begin{abstract}
Biological systems are non-linear, include unobserved variables and the physical principles that govern their dynamics are partly unknown. This makes the characterization of their behavior very challenging. 
Notably, their activity occurs on multiple interdependent spatial and temporal scales that require linking mechanisms across scales.
To address the challenge of bridging gaps between scales, we leverage partial differential equations (PDE) discovery. PDE discovery suggests meso-scale dynamics characteristics from micro-scale data. 
In this article, we present our framework combining particle-based simulations and PDE discovery and conduct preliminary experiments to assess equation discovery in controlled settings. 
We evaluate five state-of-the-art PDE discovery methods on particle-based simulations of calcium diffusion in astrocytes. The performances of the methods are evaluated on both the form of the discovered equation and the forecasted temporal variations of calcium concentration. 
Our results show that several methods accurately recover the diffusion term, highlighting the potential of PDE discovery for capturing macroscopic dynamics in biological systems from microscopic data. 

\end{abstract}

\section{Introduction}
Biological systems are composed of numerous intertwined spatial and temporal scales, making their global behavior challenging to describe and understand. 
Computational tools such as modeling and simulation are valuable strategies to address this complexity~\cite{dada_multi-scale_2011}.

These tools allow researchers to encode biological knowledge into models, which are then simulated to reproduce the behavior of the system of interest and to conduct computational studies. The resulting synthetic data are then compared to real-world observations, providing useful information and predictions on the simulated system. 
Considering that biological objects are high-dimensional and complex, the field benefits from the recent advances in Artificial Intelligence. This trend is part of a broader movement known as ``AI for science'', where machine learning and data-driven methods are used to accelerate scientific discovery and modeling.

Here, we study molecular diffusion in the cytosol of a cell, which approximates a signal propagating within the cell. 
Even though various molecular interactions are well understood, their behavior at the cell scale often remains mysterious. The complex interactions between entities may lead to non-linear, unexpected, emergent behaviors that may be observed at the macro-scale.
In addition, at the microscopic scale, reactions often occur below the diffraction limit, which challenges studying them in live tissue~\cite{fuhrmann_super_2022}. As a result, models are often fitted and validated with observations conducted at the macro-scale level. 

A key challenge in this context is the transition from smaller to larger scales, in our case from microscopic to macroscopic dynamics.  
We face a problem where, while we are able to describe micro-scale mechanisms, assessing the model against macro-scale observations requires simulating the biological phenomena at the macro-scale. 

Bridging this gap between scales would enable scientific progress, allowing modeling an entire cell from molecular interactions in cell types whose dynamics are not fully understood, such as astrocytes~\cite{verkhratsky_physiology_2018}. 
More than just validating our understanding of the mechanistic processes occurring in a cell, being able to simulate complex emerging behaviors at the macro-scale will enable to conduct prospective studies through new simulations.
Thus, it opens new possibilities for understanding complex biological systems in an interpretable and scalable way, even when detailed mechanistic knowledge is inaccessible.

In this work, we propose to leverage PDE discovery to bridge the scale gap. 
Partial Differential Equations (PDE) have often been used as tools to model spatially and temporally biological processes. 
However, implementing such models requires a deep understanding of the mechanisms and is time-consuming or impractical for complex systems.
This direction aligns with the increasing interest in applying PDE discovery to biological systems, as presented in recent work~\cite{brunton_promising_2024}, which highlights the need for innovation to address challenges such as non-stationarity or spatial heterogeneity.
PDE discovery offers a promising alternative by learning PDEs directly from experimental or, in our case, synthetic data.  
Intuitively, it automatizes the discovery of PDEs that fit the observational data by efficiently exploring a space of possible modeling of the system dynamics. 
As most recent machine learning approaches, it requires a lot of data. 
In our framework, these data are generated by a mechanistic model at the micro-scale.

In the remainder of the article, we start by introducing standard approaches to model biological processes and state-of-the-art PDE discovery tools.
Then, Section~\ref{sec:framework} formalizes the general approach of our framework. 
A preliminary version of this framework has been implemented to tackle the modeling of calcium ions diffusion in astrocytes. 
This biological context is introduced to motivate the overall approach and evaluate its potential. 
A key of its success is the effectiveness of PDE discovery methods to reveal the true PDE. 
A simplified mechanistic model of calcium diffusion based on Brownian motion is used to carry out a benchmark of different PDE discovery methods and different conditions.\footnote{We propose this biological case study to make the benchmark more tangible. It can be generalized to the modeling of other diffusion-reaction processes.}
Section~\ref{sec:results} presents the results of this benchmark.

\section{Related Works}\label{sec:soa}
\subsection{Modeling in Biology}
Deterministic models, such as Ordinary Differential Equations (ODE) and Partial Differential Equations (PDE), are classical approaches that are used to describe continuous biological phenomena~\cite{bachar_introduction_2013}. These models are based on mechanistic assumptions about processes involving diffusion, reaction, transport, and provide a compact representation of the system dynamics. Tools such as COPASI~\cite{hoops_copasicomplex_2006} for ODE models, or BioNetGen~\cite{harris_bionetgen_2016} for rule-based modeling (which can generate ODEs when applicable), are commonly used to simulate biochemical reactions. The construction of such models requires detailed knowledge of the underlying mechanisms and a careful selection of terms and parameters. However, identifying the laws responsible for the dynamics of interest is not always straightforward. 
In addition, deterministic models do not account for random fluctuations or uncertainties~\cite{hahl_comparison_2016}, which can be important in biological systems. 

Particle or voxel-based approaches are commonly used to simulate the spatio-temporal dynamics of stochastic processes. They simulate the dynamics of individual entities (molecules, cells, etc.) following mechanistic rules and allow the analysis of the global behavior of the system. 
Existing simulators include Smoldyn~\cite{andrews_detailed_2010} and MCell~\cite{kerr_fast_2008} (particle-based) or STEPS~\cite{hepburn_steps_2012} (voxel-based). These models are well-suited to capture spatial heterogeneity and local interactions but are computationally intensive, especially as the number of entities or the domain size increases. 
Extracting a macroscopic description or general laws from such simulations remains a significant challenge.

Hybrid models, which couple particle-based simulations with PDE or ODE frameworks offer a powerful strategy for capturing both microscopic stochasticity and macroscopic dynamics in cell biology. Smith et al.~\cite{Smith2018}  reviewed spatially-extended hybrid methods to model multiscale biological and physical systems. 

\subsection{PDE Discovery}

ODE/PDE discovery is an emerging machine learning technique to identify governing laws from experimental data. 
This field has advanced in parallel with sparse regression, machine learning, and deep learning. 
These methods have yielded promising results in various biological contexts, such as  sea surface height dynamics~\cite{maslyaev_partial_2021} or cell migration and proliferation~\cite{chen_physics-informed_2021} modeling. 
This section briefly reviews PDE discovery methods by type and chronology.

PDE discovery is usually presented formally as the problem of discovering a (non-linear) function $F$ such that:
\begin{equation}\label{eq:PDE}
\vec{u}_t = F(\vec{u}, \vec{u}_x, \vec{u}_{xx}, \dots, \vec{u}^2_x, \vec{u}^2_{xx}, \dots , x),
\end{equation}
where $\vec{u}_t=\frac{\partial \vec{u}}{\partial t}$ is the first-order time derivative of the field $\vec{u}(t,x)$, and $\vec{u}_x\dots \vec{u}_{xx}$ are the spatial derivatives of $\vec{u}$.\footnote{For the sake of simplicity, this article considers only one spatial dimension $x$ but all the methods can handle more spatial dimensions.} 

Early PDE discovery methods rely on the assumption that most equations can be expressed as a sparse linear combination of terms from a predefined set ($\vec{u}_x$, $\vec{u}_{xx}$, etc.), called a library. They make the Ockham's razor assumption and avoid more complex PDE.
This approach is used in SINDy~\cite{brunton_discovering_2016} and PDE-Find~\cite{rudy_data-driven_2017}, which perform sparse regression to select the relevant terms from this fixed library to get a PDE. 
However, if the underlying dynamics is driven by a model that is out of this dictionary, it cannot be discovered. To address this, more flexible methods such as SGA-PDE~\cite{chen_symbolic_2022} and DLGA-PDE~\cite{xu_dlga-pde_2020} use genetic algorithms to generate custom libraries, expanding the search space at the cost of increased computational cost. 
Another challenge for sparse regression methods is the sensitivity to noise, as estimating derivatives from noisy data can degrade accuracy. To address this, weak formulations integrate PDEs over domains with test functions. This smooths the derivatives and improves the robustness. 
Weak SINDy~\cite{messenger_weak_2021} combines the use of weak formulations with sparse regression to handle noise.

Despite improvements, symbolic methods often rely on predefined structures and struggle in high-dimensional settings. Deep learning offers more flexible alternatives. Convolutional Neural Network (CNN) based models such as PDE-Net~\cite{long_pde-net_2018,long_pde-net_2019} aim to simultaneously learn the solution and the underlying PDE structure. DeepMoD~\cite{both_deepmod_2021} approximates both the target and its derivatives through automatic differentiation~\cite{baydin_automatic_2018}, minimizing the effects of noise during sparse regression. These models are powerful, but more computationally demanding. Hybrid methods combine the interpretability of physics-based methods while incorporating the adaptability of machine learning, leading to more robust and generalizable solutions. 

An alternative to differentiation to simulate a PDE is to use Physics-Informed Neural Networks (PINNs)~\cite{raissi_physics-informed_2019}. 
PINNs integrate PDE constraints into the loss function of neural networks, ensuring that the discovered solutions respect the underlying physics of the system. R-DLGA~\cite{xu_robust_2021} constructs a library of candidate terms using a neural network and a genetic algorithm. Then, a PINN is used to discover potential terms which are added to the loss function as physical constraints to further optimize the derivatives and discover the PDE. 

In parallel, probabilistic methods that provide predictions while accounting for uncertainty have emerged. Instead of assuming that there is only one correct equation that successfully describes the system dynamics, these methods recognize that various equations might fit the data, especially when the data is noisy, incomplete, or complex~\cite{pergler_probabilistic_2008}. For example, Bayesian Symbolic Learning (BSL)~\cite{sun_bayesian_2022} leverage Bayesian inference to account for uncertainty in noisy datasets, improving the robustness of discovered equations.

Finally, Reinforcement Learning (RL) has been used to explore complex and high-dimensional solution spaces. For example, R-DISCOVER~\cite{du_physics-constrained_2024} applies physics-constrained reinforcement learning to identify governing equations, while MORL4PDEs~\cite{zhang_morl4pdes_2024} deploys multi-objective RL to balance accuracy, simplicity, and efficiency.

We also make an Ockham's razor assumption for all of the tested PDE discovery methods in this study assuming that the underlying equation can be expressed as a linear combination of candidate terms (see section~\ref{sec:protocol}), as this formulation allows the use of efficient sparse regression techniques to identify the governing dynamics.

\section{Combining Scales through PDE Discovery}\label{sec:framework} 

\begin{figure}[tb]
\centering
\includegraphics[width=.7\textwidth]{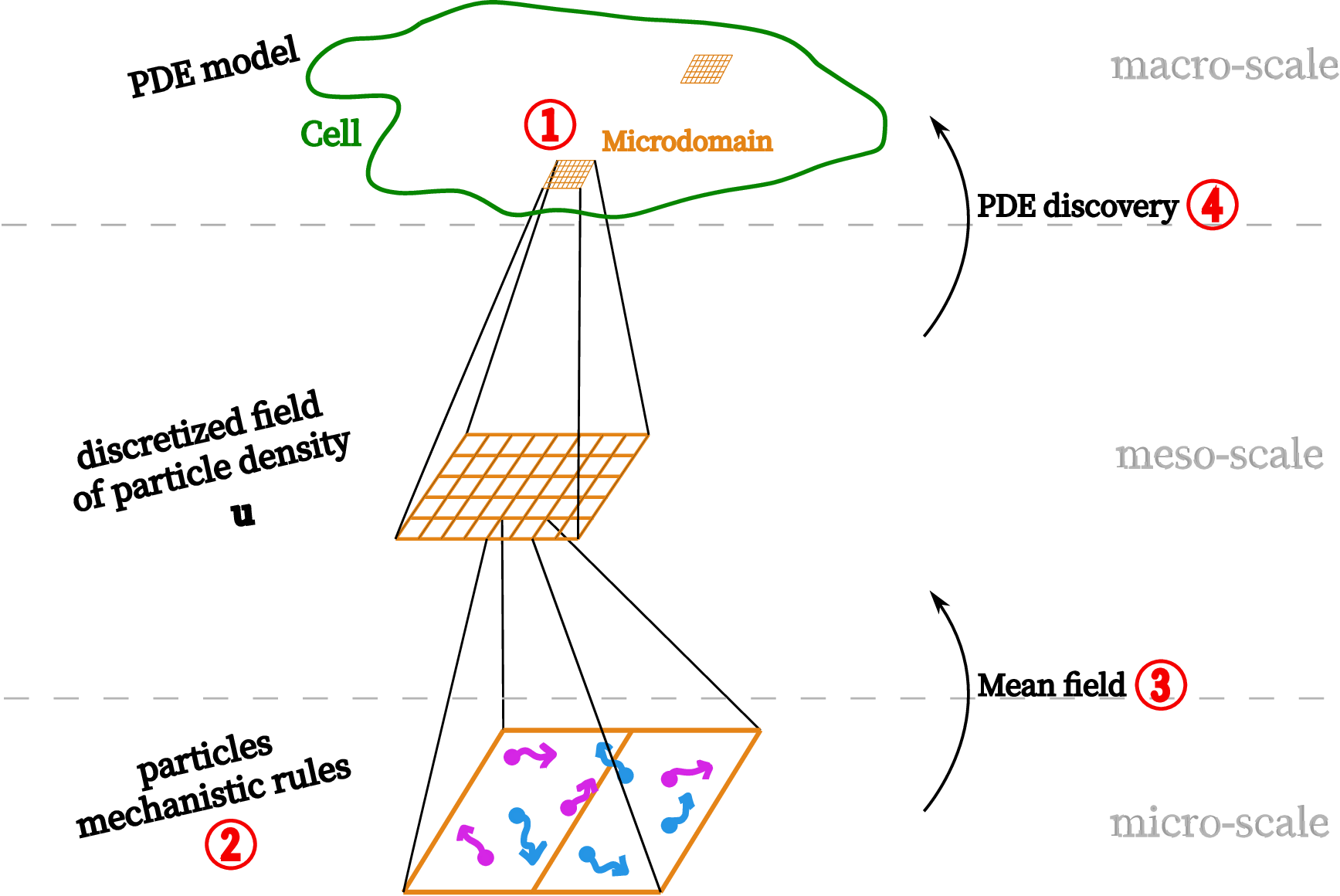}
\caption{Framework of the proposed approach for macro-scale simulations using PDEs discovered from micro-scale simulated data. 
Briefly, the diffusion of calcium ions in sub-cellular compartments referred to as microdomains (1) are simulated numerically (2). Then, local mean fields of particle density are computed at the meso-scale (3), on which PDE discovery methods are applied to infer the equation governing the macro-scale (4).}
\label{fig:framework}
\end{figure}

Figure~\ref{fig:framework} illustrates the overall approach of this study. We propose to simulate macro-scale processes from micro-scale mechanistic rules. 
The micro-scale process is the evolution of the density of some particles represented by a spatio-temporal field $\vec{u}(t,x)\in\mathbb{R}_+^d$, where $d$ represents the dimension of the field (e.g. the number of different types of particles). 
The meso-scale corresponds to a spatio-temporal grid used to simulate a PDE. Because of the assumptions of the state-of-the art PDE discovery methods, the grid is regular. 
We also assume that the boundary conditions of the simulation are known.

The overall process consists of five main steps:
\begin{enumerate}
\item \textbf{Generate diverse realistic initial conditions.} In this first step, different spatial patches with their local boundary conditions are randomly sampled in the whole macro-system to simulate (e.g., the whole cell).
\item \textbf{Simulate mechanistic local behaviors for each patch}. Then, a mechanistic particle-based engine locally simulates the microscopic behavior of the system. This spatio-temporal range of the simulation is set to balance the computational cost of the simulation and the need for data of PDE-discovery methods.
\item \textbf{Mean fields of the local simulation.} Along the simulations, the density of particles at the meso-scale are computed to create a dataset representing the local behavior of the system. This step creates $n$ local mean fields $(\vec{u}_i)_{i\in[n]}$. Note that several runs of the same patch can yield different field values due to the stochasticity of the simulation. 
\item \textbf{PDE discovery.} A PDE discovery method is applied to the collection of spatio-temporal fields that have been simulated to infer the governing equation, represented by the $F$ function. This process assumes that there is only one governing equation for the entire macro-system. 
\item \textbf{Simulate the global behavior.} The discovered differential equation is integrated numerically on the entire spatio-temporal domain to simulate the macro-scale behavior.
\end{enumerate}

It is worth noticing that the proposed method is data-free. 
This model only requires the knowledge of mechanistic rules to generate as much data as necessary to robustly discover the governing PDE. 
Nonetheless, if some observational data have been collected, they can be integrated in the pipeline as additional mean fields to the synthetic ones. Thus, the overall process can be seen as a way to augment the dataset for PDE-discovery.

\section{Evaluation of the Framework for Calcium Signaling Characterisation in Astrocytes}\label{sec:usecase}

The overall principle presented in the previous section assumes that PDE discovery methods are able to accurately identify the correct equation from generated data. 
In our framework, we evaluate the ability of these methods not only to accurately simulate the process but also to discover an equation that applies to different simulations. 

The objective of this use case is to investigate, on a realistic but controlled dataset, whether the state-of-the-art PDE discovery methods can achieve this objective of equation discovery accurately. 
The scenario is said to be \textit{realistic} because it is inspired by a biological question, presented in section~\ref{sec:biological_context}. 
It is said to be \textit{controlled} because we implemented a simple mechanistic model whose mean field behavior is a known PDE.

\subsection{Application Context}\label{sec:biological_context}
We investigate the use of PDE discovery methods to study calcium signaling in astrocytes. Astrocytes are the most abundant non-neuronal cells in the central nervous system, and play a key role in modulating neuronal activity through calcium signaling~\cite{verkhratsky_physiology_2018}.

Most calcium signals in astrocytes, notably those associated with neuron–astrocyte communication, occur in microdomains and are spatially restricted~\cite{ahrens_astrocyte_2024}. These signals are difficult to investigate experimentally because they occur in compartments below the diffraction limit, which is why particle-based models have been developed to study them~\cite{denizot_simulation_2019}.

Studying whole-cell dynamics is essential to understand how signals interact and integrate within an astrocyte. This is particularly relevant given that a single astrocyte can interact with more than 100,000 synapses simultaneously\cite{bushong_protoplasmic_2002}, generating numerous microdomains in the cell. 

While particle-based simulations are ideal for capturing sub-diffraction calcium dynamics, they are computationally demanding, which makes it impractical to simulate calcium signaling at the scale of the whole cell.

To bridge this gap, we propose to test whether PDE discovery methods could be used to simulate calcium dynamics in astrocytes at the whole-cell level. More precisely, we generated a synthetic dataset using particle-based simulations of calcium diffusion in confined spaces. 
This simulation is based on a Brownian motion of particles. 

Our aim is to evaluate whether the existing PDE discovery methods can reconstruct the correct governing equations. We know how to describe several biological reactions with mechanistic modeling such as diffusion processes, which can be described by Einstein's relation for Brownian motion~\cite{einstein_investigations_1926}. 
Here, the expected differential equation for our system is a classical diffusion equation:
$$\vec{u}_t = D \vec{u}_{xx},$$ 
where $D\in\mathbb{R}_+$ is the diffusion coefficient, and $\vec{u}$ is 
a scalar field representing the density of a unique particle. 
$D$ is a constant describing how fast particles can spread, which depends on the nature of these particles and on the properties of the system such as the temperature.

\subsection{Experimental Protocol}\label{sec:protocol}

\begin{figure}[tb]
\centering
\includegraphics[width=.31\textwidth]{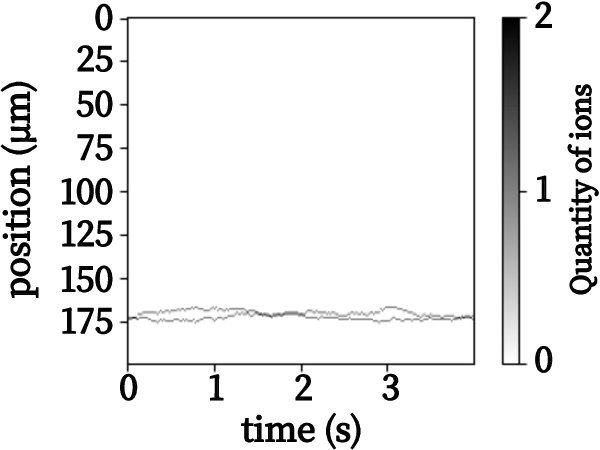}
\hfill
\includegraphics[width=.3\textwidth]{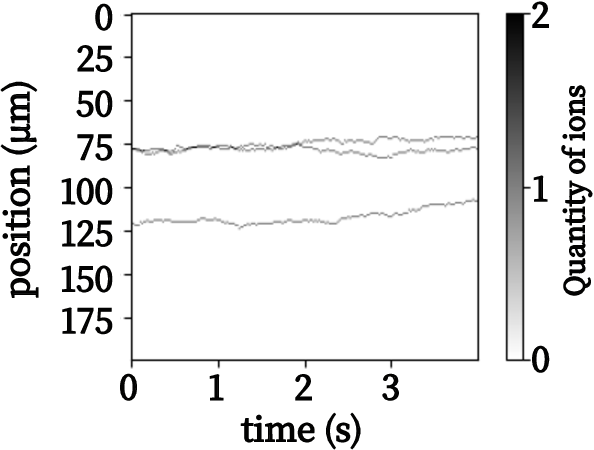}
\hfill
\includegraphics[width=.3\textwidth]{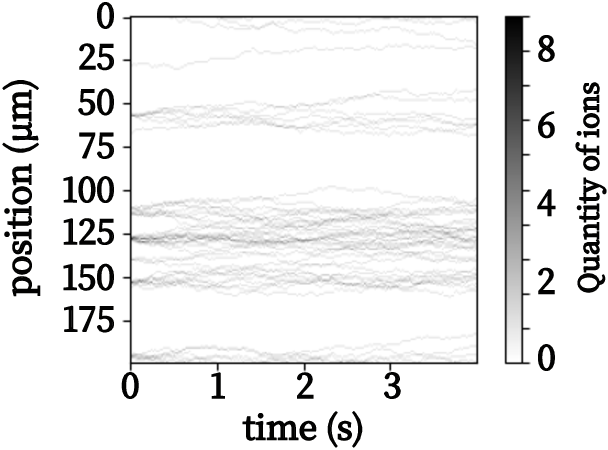}
\caption{Examples of meso-scale calcium diffusion data generated with different initial parameter values. Left: 1 site and 2 initial ions, middle: 2 sites and 3 initial ions, right: 10 sites and 50 initial ions.}
\label{fig:example_data}
\end{figure}

The data used in this work simulates calcium diffusion in a one-dimensional space, capturing the temporal evolution of calcium ions coordinates. 
This data is produced using a particle-based simulator, which models calcium ions as individual agents diffusing in space with a diffusion coefficient~$D$. Agents are initially placed at predefined locations (called \textit{sites}), and their movements are tracked over time. A function is reconstructed at each timestep by discretizing space and counting the number of particles per slice of space. This generates a spatiotemporal field $\vec{u}(x,t)$ which is used as an input to the PDE discovery methods. Figure~\ref{fig:example_data} illustrates representative generated data.

We emphasize that the objective of these experiments is not to make biological findings, but to assess the effectiveness of our framework to recover the governing dynamics from microscopic data. The coefficient of diffusion of calcium is set to D=13 $\mu m^2.s^{-1}$ to account for calcium buffering in cells~\cite{allbritton1992range}. Despite our attempt to make the synthetic data realistic, the goal of this study is not to yield the richness and complexity of state-of-the-art models of calcium signaling in astrocytes.

\vspace{10pt}

\begin{figure}[tb]
    \centering
    \includegraphics[width=0.85\linewidth]{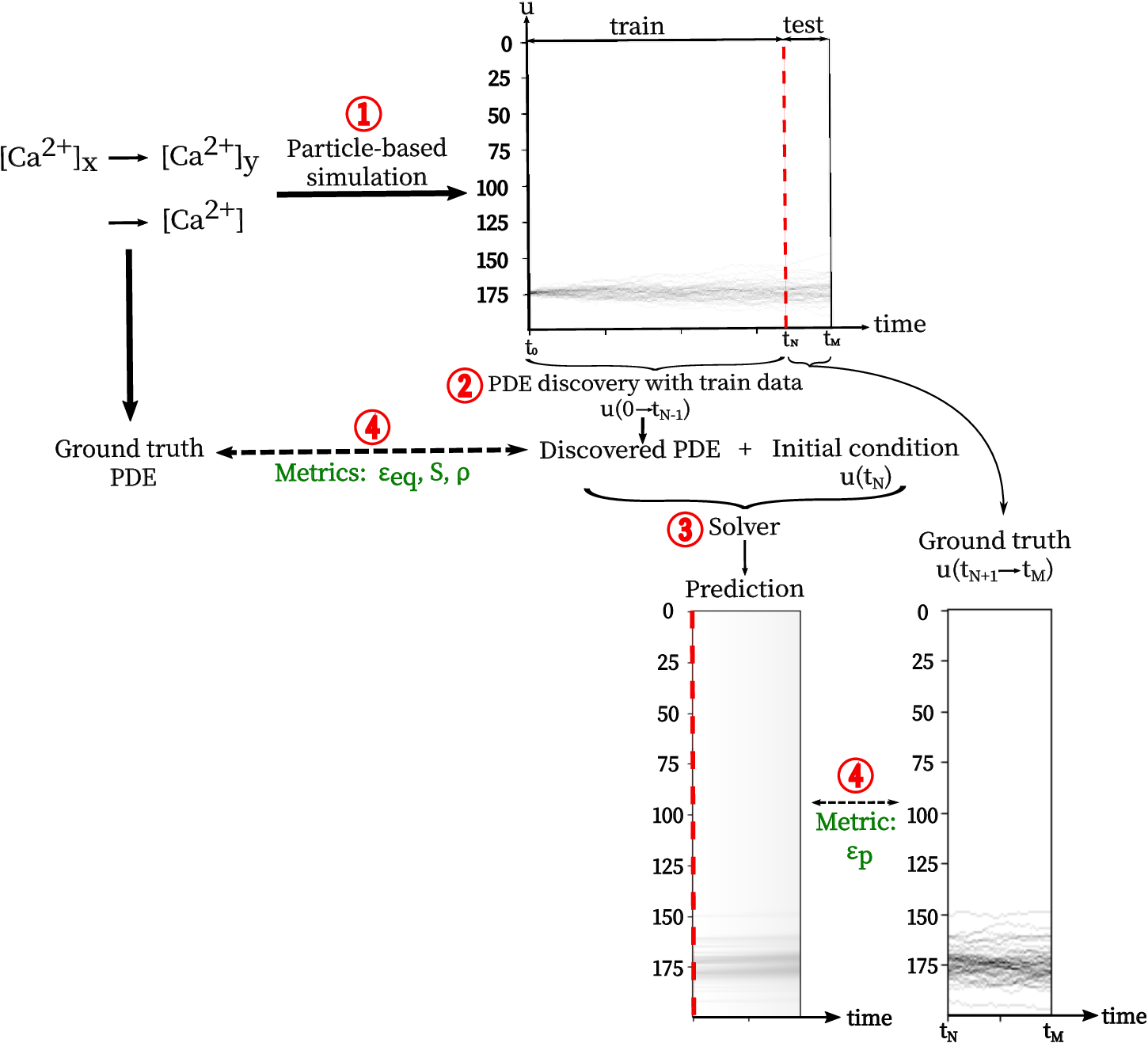}
    \caption{Framework to evaluate the performance of a PDE discovery method. \textcircled{1} generation of particle-based calcium diffusion data, \textcircled{2} PDE discovery predicting an equation, \textcircled{3} integration of the discovered equation into a numerical solver, \textcircled{4} performance evaluation using different metrics: $\varepsilon_{eq}$, $\varepsilon_p$, $S$, and $\rho$.}
    \label{fig:benchmark_framework}
\end{figure}

PDE discovery methods are evaluated using the framework presented in figure~\ref{fig:benchmark_framework}. 
The simulated data is divided into two subsets along the temporal dimension: the first 200 time steps (80\%) have been designated for training and the remaining 50 last (20\%) reserved for testing. To understand the conditions under which each method performs optimally, the entire process is repeated across a wide range of biologically relevant initial conditions.\footnote{Each set of initial conditions generates a unique field dataset used for PDE-discovery.} Specifically, the initial number of agents and sites ranged respectively from 2 to 100 and from 1 to 100 in steps of 5. Ten simulations were run for each of these initial conditions. This approach enables the identification of scenarios in which specific methods exhibit superior performance.

We selected five different methods: PDE-Find~\cite{rudy_data-driven_2017}, SGA-PDE~\cite{chen_symbolic_2022}, Weak SINDy~\cite{messenger_weak_2021}, DeepMod~\cite{both_deepmod_2021}, D-CIPHER~\cite{kacprzyk_d-cipher_2023}, which all extract the $F$ function of the PDE in eq.~\ref{eq:PDE} as a linear combination of derivative terms $f$ in a finite dictionary $\mathcal{D}$, i.e.:
\begin{equation}\label{eq:disc_PDE}
\vec{u}_t = \sum_{f\in\mathcal{D}}\widehat{\xi}_{f} f.
\end{equation}

Additionally, we designed a naive method, called \textit{Caterpillar}, specifically for the prediction task. This method serves as a baseline to discuss the results of the other approaches. Rather than discovering an equation, Caterpillar simply propagates the last observed state from the training data for all future time steps. Consequently, it is included only in the evaluation of the prediction error ($\varepsilon_p$).

Unfortunately, the collection of derivative terms depends on the method. In order to make a fair comparison between them, we propose a unified dictionary $\mathcal{D}$ of derivative terms. 
Let $p\in \mathbb{N}^p$, and $\vec{\gamma} \in 
[\![0,p]\!]^p$ which represents a vector of size $p$ containing natural integers between $0$ and $p$. An element $f_{\vec{\gamma}}\in\mathcal{D}$ is defined by:
$$f_{\vec{\gamma}} = \prod_{i\in[p]}\frac{\partial^i \vec{u}^{\vec{\gamma}_i}}{\partial x^{i}}.$$

For instance, with $p=2$, the equation $3.2 \vec{u}\vec{u}_x + 2 \vec{u}_{xx}$  is represented by coefficients $\xi_{[1,1,0]}=3.2$ and $\xi_{[0,0,1]}=2$, all other coefficients being zero.
All benchmarked methods have been set up and adapted to yield equations aligned with a dictionary $\mathcal{D}$ with $p=3$. 
$\widehat{\xi}_{\vec{\gamma}}$ denotes the fitted parameters of a PDE expressed in this dictionary and $\xi_{\vec{\gamma}}$ its true parameters.

Rather than limiting the assessment of PDE discovery performance to prediction accuracy ($\epsilon_p$), we also evaluate the accuracy of the recovered equation using three metrics ($\epsilon_{eq}$, $S$, $\rho$):
\begin{itemize}
\item $\varepsilon_{eq}$ calculates the Mean Squared Error (MSE) between the discovered PDE and the ground-truth equation. Low MSE values indicate an accurate representation of the system dynamics. As we evaluate the capacity of the methods to discover the diffusion equation, the absence of the diffusion term is penalized.
$$
\varepsilon_{eq} = \left\{\begin{array}{ll}
    \infty & \text{if } \xi_{[0,0,1,0]}=0 \text{ (no diffusion term)} \\
     \sum_{\vec{\gamma}} (\xi_{\vec{\gamma}} - \widehat{\xi}_{\vec{\gamma}})^2& \text{otherwise} \\
    \end{array}\right.
$$

\item $\varepsilon_{p}$ calculates the Mean Squared Error (MSE) between the forecasted (from the discovered PDE) and the ground-truth next states of $\vec{u}$.
$$\varepsilon_{p} = ||\vec{u}(t_N\dots t_M,\cdot) - \hat{\vec{u}}(t_N\dots t_M,\cdot)||^2,$$
where $\vec{u}(t_N\dots t_M,\cdot)$ refers to the true numerical values of test data, and $\hat{\vec{u}}(t_N\dots t_M,\cdot)$ represents values of predicted data (see Fig.~\ref{fig:benchmark_framework}).
    
    \item $S$ measures the sparsity of the discovered equation. 
$$
S = \left\{\begin{array}{ll}
    \infty & \text{if } \xi_{\vec{\gamma}}=0,\; \forall \vec{\gamma} \\
    1-\frac{\left|\left\{\xi_{\vec{\gamma}}\mid \xi_{\vec{\gamma}}=0\right\}\right|}{\left|\left\{\xi_{\vec{\gamma}}\right\}\right|} & \text{otherwise} \\
    \end{array}\right.
$$
$S$ is close to 1 for a more compact and interpretable model, which is desirable to identify the dynamics of the system.

    \item $\rho$ is the rank of the diffusion term, where terms are listed by descending order of their coefficient value, i.e. from the most important to the least important. Comparing the ranks of terms in the true equation with that of the discovered equation provides an insight into the ability of the method to correctly prioritize the key terms within the equation.
\end{itemize}

\subsection{Benchmark results}\label{sec:results}
In this section, we present the performance of the PDE discovery methods within our modeling framework.\footnote{Our implementation of the framework, including the generator of synthetic datasets, is accessible for reproducibility purposes: \url{https://gitlab.inria.fr/tguyet/pde-benchmark.git}.} 

\begin{figure}[!h]
 \centering
 \begin{subfigure}[h]{0.49\textwidth}
     \centering
     \caption{$log(\varepsilon_{p})$}
     \includegraphics[width=\textwidth]{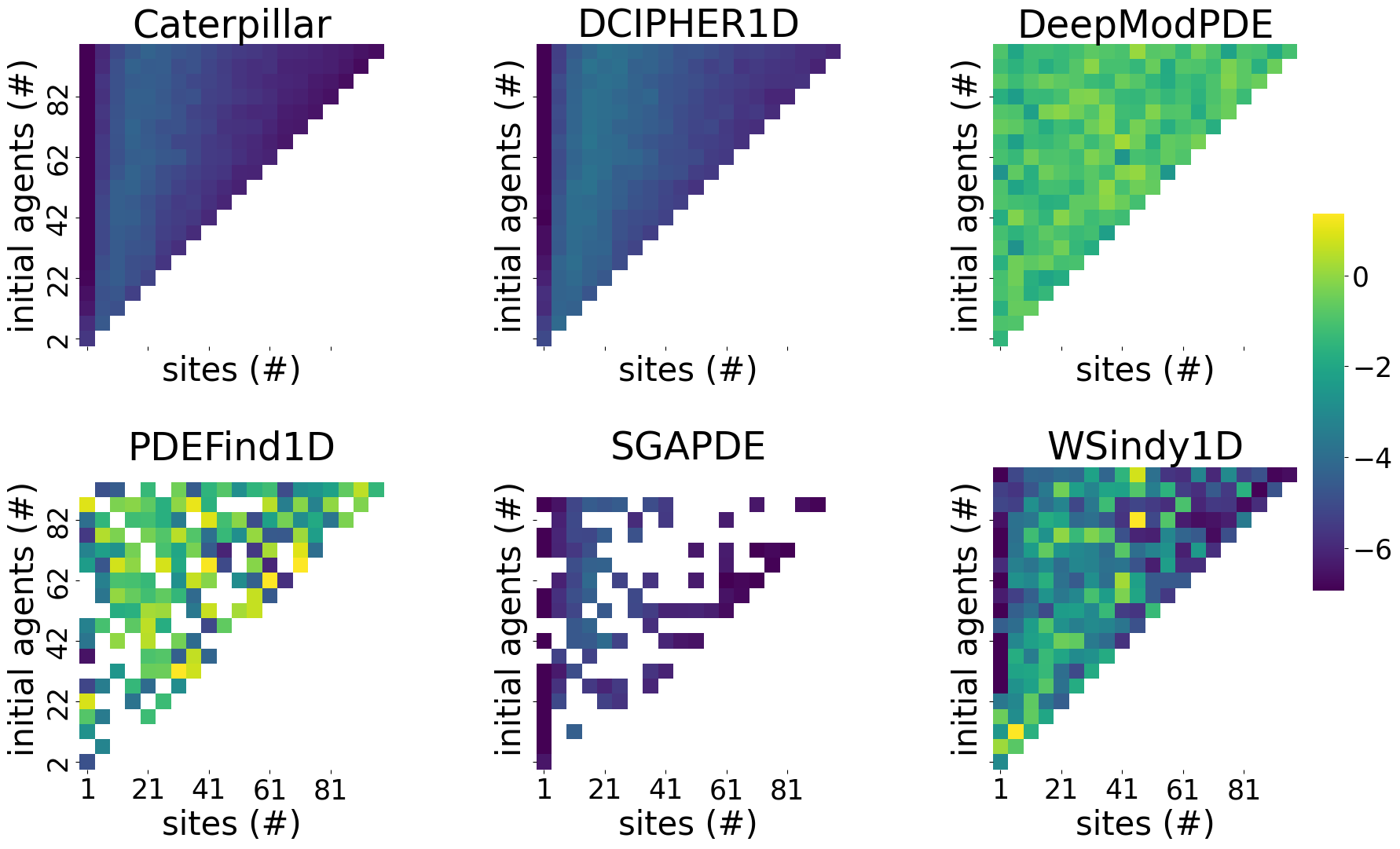}
     \label{fig:results_predmse}
 \end{subfigure}
 \hfill
 \begin{subfigure}[h]{0.49\textwidth}
     \centering
     \caption{$log(\varepsilon_{eq})$}
     \includegraphics[width=\textwidth]{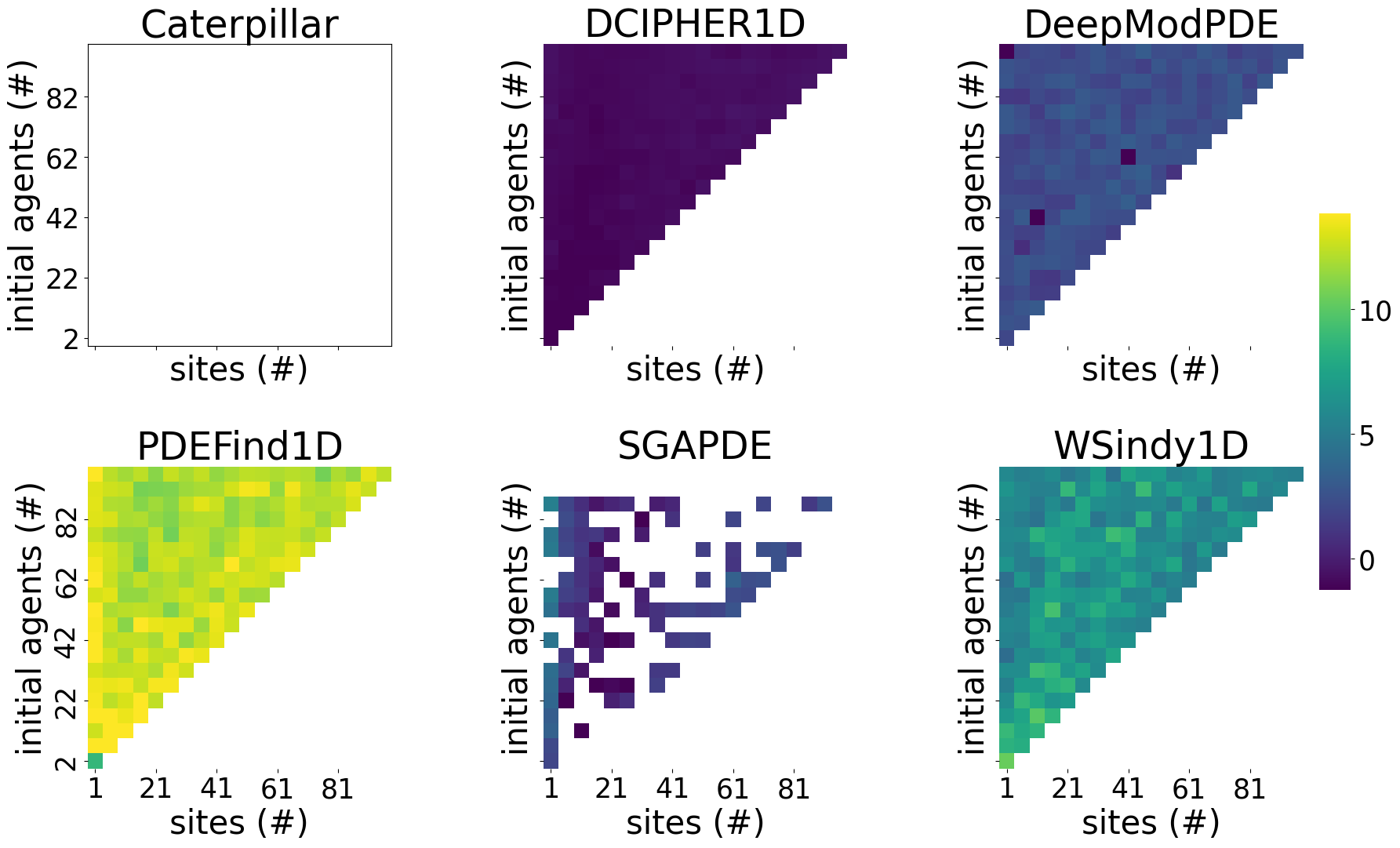}
     \label{fig:results_eqmse}
 \end{subfigure}
 \begin{subfigure}[h]{0.49\textwidth}
     \centering
     \caption{$S$}
     \includegraphics[width=\textwidth]{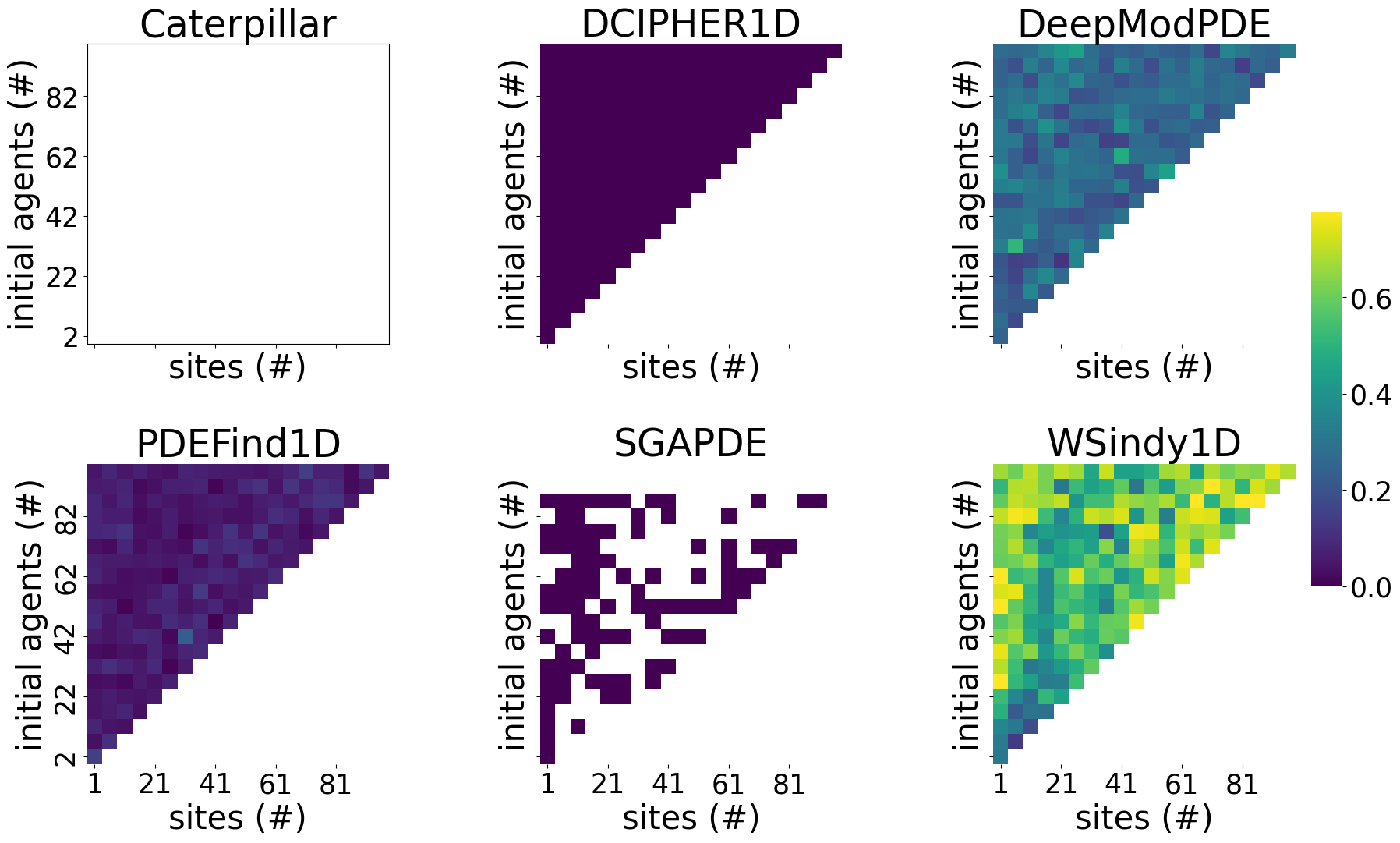} 
     \label{fig:results_sparsity}
 \end{subfigure}
 \hfill
 \begin{subfigure}[h]{0.49\textwidth}
     \centering
     \caption{$\rho$}
     \includegraphics[width=\textwidth]{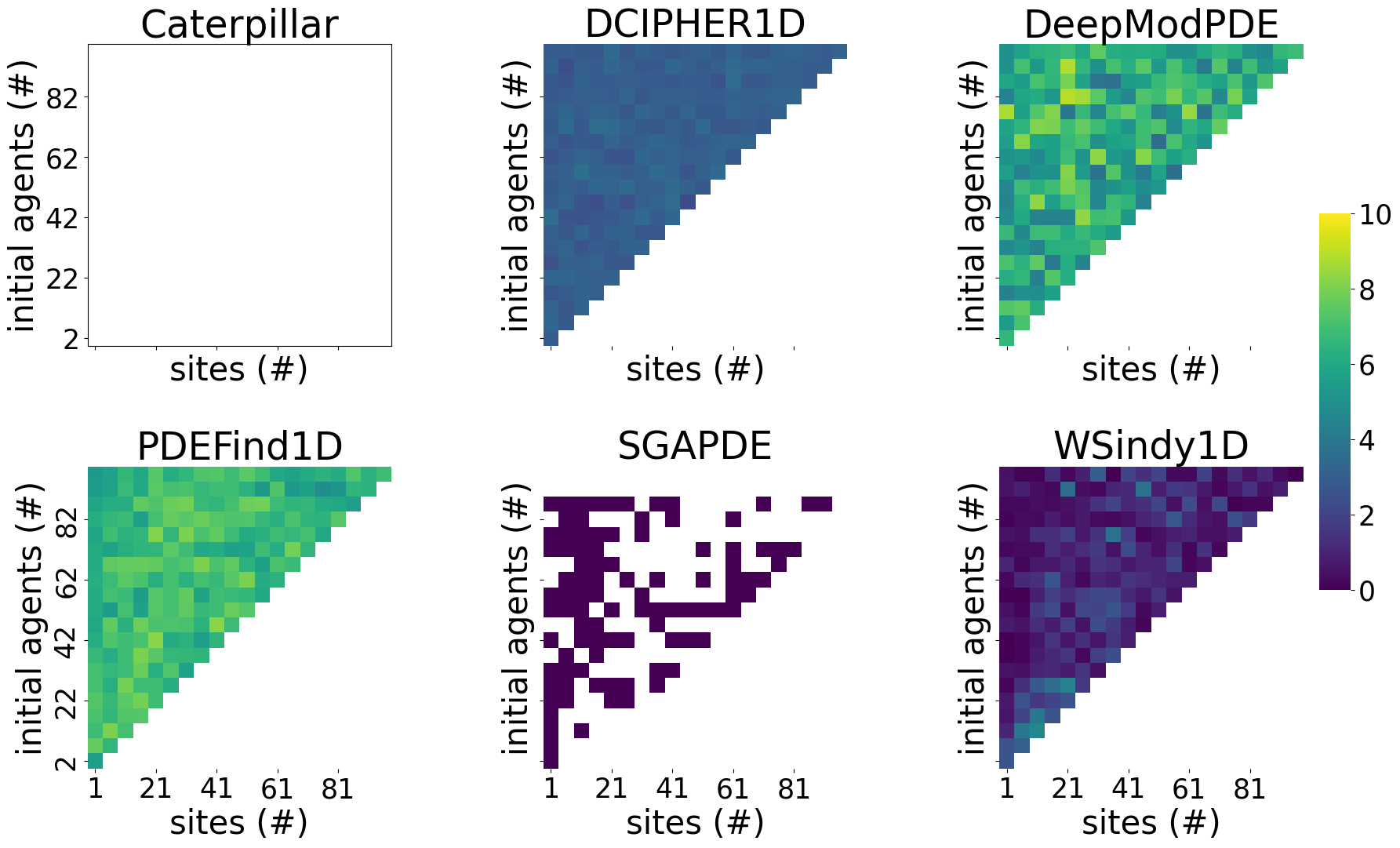}
     \label{fig:results_order_coeff}
 \end{subfigure}
    \caption{Heatmaps of method performance across varying numbers of initial agents and sites for four evaluation metrics. Each heatmap displays the average value over experiments with the same initialization of a given metric as a function of the number of initial agents (rows) and initial sites (columns). The four metrics are: 
(a): $\varepsilon_{p}$ measures the MSE between the ground truth and forecasted next states of $\vec{u}$. Darker blue regions correspond to lower errors and indicate a better performance. 
(b): $\varepsilon_{eq}$ measures the MSE between the discovered and true equation. Darker blue regions correspond to lower errors and indicate a better performance. 
(c): $S$ indicates the level of sparsity of the discovered equation. Lighter regions correspond to a sparsity close to 1 and indicate a better performance. 
(d): $\rho$ evaluates the diffusion term's rank. Darker blue regions correspond to a better prioritization of the key terms within the equation and indicate a better performance.}
    \label{fig:metrics_diffusion}
\end{figure}

Figure~\ref{fig:metrics_diffusion} illustrates the performance of the PDE discovery methods according to the different metrics for each setting (number of agents from $2$ to $100$, number of sites from $1$ to $100$).

The empty cells in figure~\ref{fig:metrics_diffusion} correspond to conditions where a PDE discovery method failed to yield results. These failure cases will be detailed in the following paragraphs.

Regarding the prediction error $\varepsilon_p$ (Fig.~\ref{fig:results_predmse}), D-CIPHER and Caterpillar are overall the most effective methods (lower error). 
WSINDy shows good performance where many sites and agents are present, suggesting its strength in dense environments. 
Caterpillar, D-CIPHER, and SGA-PDE exhibit similar behaviors, mainly extending the trajectories from their initial conditions (i.e., the last configuration of the training data). It is difficult to draw any conclusions about PDE-Find and DeepMoD because they show mixed results with variable performance across different configurations.
D-CIPHER, DeepMoD and SGA-PDE appear to be the most effective methods for finding the true equation (Fig.~\ref{fig:results_eqmse}). 
It is interesting to note that even if DeepMod finds an equation rather close to the diffusion equation, the forecast was not as good as the Caterpillar method. In contrast, even if WSINDy did not find the right equation, it yields an accurate prediction (in dense settings). This illustrates the need to evaluate both the quality of the prediction and the quality of the discovered equations. 
As the $\varepsilon_{eq}$ metric only measures the distance between the ground-truth and the discovered equation, methods that predict small coefficients for various terms can still achieve a low error on the equation ($\varepsilon_{eq}$) scores. Therefore, it is essential to also consider the sparsity metric $S$.
PDE-Find rarely produces sparse equations, which can explain its good performance in $\varepsilon_{eq}$ (Fig.~\ref{fig:results_sparsity}). In contrast, Weak SINDy achieves the best results for sparsity. Notably, D-CIPHER and SGA-PDE never yield equations with zero coefficients.
Finally, Weak SINDy and D-CIPHER perform better than the other methods on the order coefficient metric $\rho$ (Fig.~\ref{fig:results_order_coeff}), while other methods generate comparable results.

\begin{figure}[tb]
 \centering
 \begin{subfigure}[h]{0.95\textwidth}
     \centering
     \caption{Number of excluded experiments (no diffusion term)}
     \includegraphics[width=\textwidth]{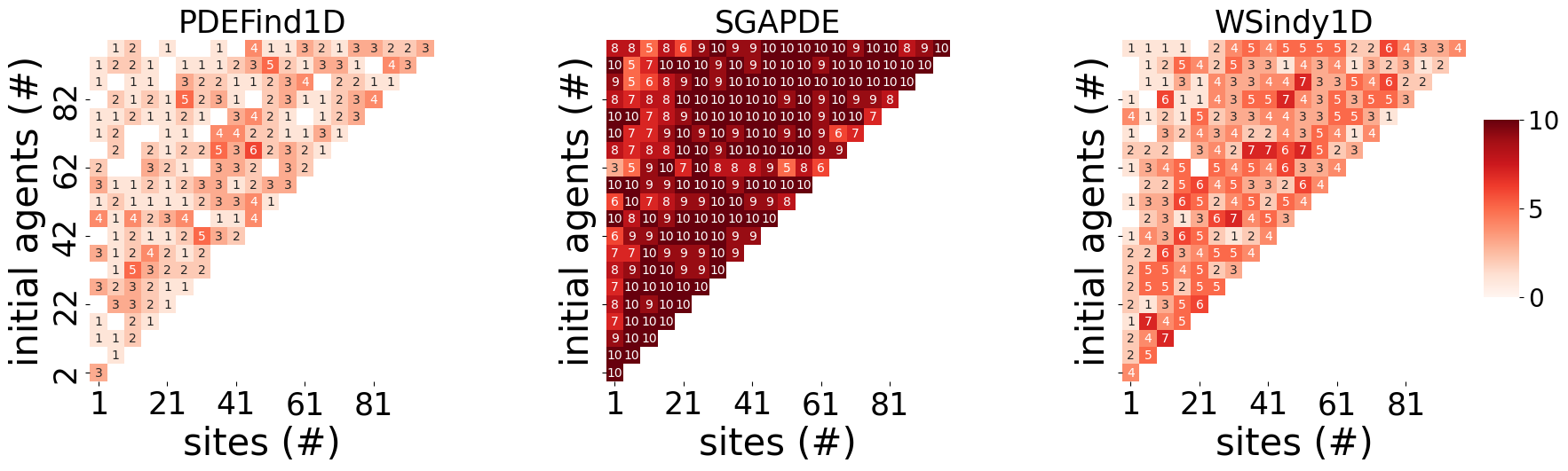}
     \label{fig:exclu}
 \end{subfigure}
 \hfill
 \begin{subfigure}[h]{0.6\textwidth}
     \centering
     \caption{Number of \texttt{NaN} (numerically unstable discovered equation)}
     \includegraphics[width=\textwidth]{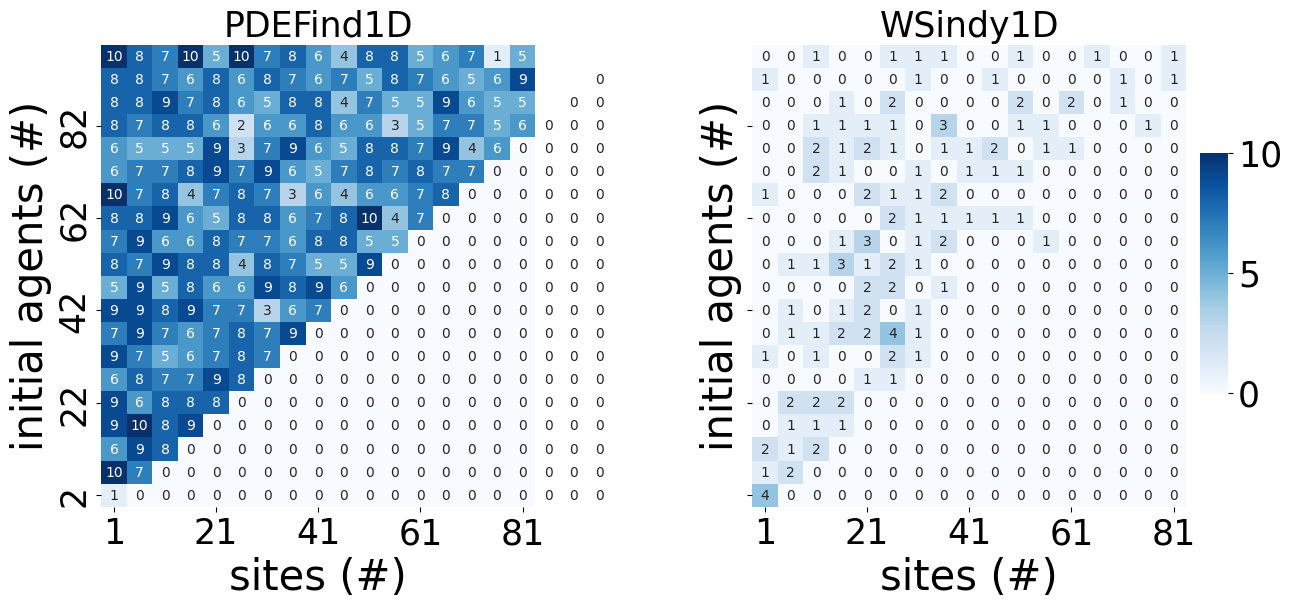}
     \label{fig:nan}
 \end{subfigure}
    \caption{Excluded cases and NaN occurrences across experimental settings. (a) Number of excluded values per configuration, for three methods (PDEFind1D, SGAPDE, WSindy1D). (b) Number of NaN values for the $\varepsilon_p$ metric in two methods (PDEFind1D and WSindy1D), corresponding to scenarios where the solver failed to integrate the discovered equation and to forecast future values of $\vec{u}$. These visualizations highlight which experimental conditions are most subject to failures or incomplete discovery. Ten simulations were run for each set of initial conditions.}
    \label{fig:exclunan}
\end{figure}

Two different failure situations can be observed. 
In some cases, the equation does not even contain the diffusion term~$\vec{u}_{xx}$. In this case, the metrics have infinite values and the experiment is excluded from the results. 
Another specific case is when the equation that is discovered is too complex to be numerically integrated. Equations with many terms may lead to unstable numerical integration that yields \texttt{NaN} values during the third step of our framework (Fig~\ref{fig:benchmark_framework}). In that case, it is not possible to compute the forecast accuracy metrics $\varepsilon_p$.

Figure~\ref{fig:exclunan} illustrates the settings and the methods in which these situations have been encountered (methods that were successful in all simulated conditions are not shown). Figure~\ref{fig:nan} shows the number of simulations for which the discovered equations were too complex. This issue only arose with PDE-Find and Weak SINDy. PDE-Find shows a high frequency of solver failures. Figure~\ref{fig:exclu} shows the counts of excluded experiments. For PDE-Find and Weak SINDy, the frequency of diffusion term inclusion in the discovered PDE appears somewhat random, indicating unpredictable performance in identifying the correct physics, while SGA-PDE almost never correctly identifies diffusion as the simulated dynamics.

These results show that, even with a simple process, the PDE-discovery methods may fail to yield the true equations. A possible explanation for this observation is the small number of data provided to the machine learning methods in this framework. It is also important to note that most of the methods have been designed and experimented in a similar setting, using only one run of a dynamical process to infer the equation. 
Including several simulation runs to the pipeline will most likely improve the robustness of the method, but may result in an increased computational cost.

\begin{figure}[tb]
    \centering
    \includegraphics[width=0.43\linewidth]{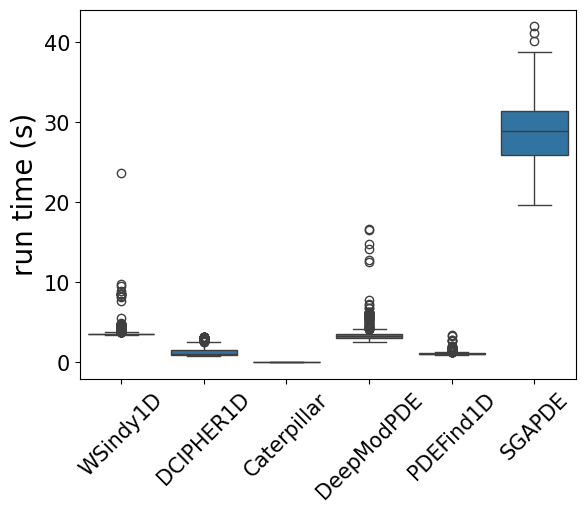}
    \caption{Running time of the benchmarked PDE discovery methods.}
    \label{fig:execution_time}
\end{figure}

For now, the computational costs of most of the methods are low. 
Figure~\ref{fig:execution_time} compares the execution time of the considered methods. 
It reveals significant differences in the efficiency of the PDE discovery methods. SGA-PDE is by far the most time-consuming approach, with high variability. In contrast, Weak SINDy and DeepMoD exhibit similar and relatively low computation times (a few seconds), making them efficient choices in terms of runtime. Caterpillar and PDE-Find are even faster.

\section{Conclusion}
We propose to study large-scale ``emerging'' phenomena using PDE discovery methods on data that is either scarce or impossible to measure experimentally, but whose microscopic dynamics is known. To support this approach, we introduce a framework that evaluates existing methods in this context. Our case study on astrocyte calcium diffusion reveals that some existing methods can predict the macroscopic dynamics from particle-based simulations. 
Among the evaluated methods, Weak SINDy stands out as a fast and effective approach with high sparsity and precise recovery of the diffusion term. D-CIPHER is also promising, though it requires further fine-tuning due to its sensitivity to the choice of candidate terms and test functions. The other methods tested in this benchmark appear less suited to our biologically-inspired particle-based data.
These differences in performance may stem in part from how each method constructs or relies on a dictionary of candidate terms: fixed and predefined for PDE-FIND and Weak SINDy; user-defined for D-CIPHER; automatically generated from data for SGA-PDE and DeepMoD. As a result, the true PDE may not be identifiable within some of these dictionaries, which can explain why certain methods yield good predictions without correctly recovering the equation sparsity.
These first results need to be confirmed with controlled synthetic data generated by more complex reactions involving several molecules, in 2D/3D spatial domains, which are more realistic datasets than commonly used in the field of PDE discovery. These prospective studies, together with the development of new tools tailored to the complexity of biological data, pave the way for future research investigating the physical laws governing cell functions.

\bibliographystyle{splncs04}
\bibliography{biblio_papier_benchmark}
\end{document}